\documentclass[aps,prd,twocolumn,showpacs,preprintnumbers,nofootinbib,amsmath,amssymb]{revtex4}

\usepackage{graphicx}
\usepackage[caption=false]{subfig}
\usepackage{bm}
\usepackage{amsthm,url}
\usepackage{color,ulem}
\usepackage{cancel}
\usepackage{slashed}
\usepackage{lipsum}

\def\be{\begin{equation}}
\def\ee{\end{equation}}
\def\bea{\begin{eqnarray}}
\def\eea{\end{eqnarray}}
\def\bn{\begin{enumerate}}
\def\en{\end{enumerate}}
\def\bsube{\begin{subequations}}
\def\esube{\end{subequations}}
\def\ll{\left}
\def\rr{\right}

\def\mc{\mathcal}
\def\mb{\mathbf}
\def\nn{\nonumber}
\def\ll{\left}
\def\rr{\right}

\def\x{\mb{x}}
\def\s{\mb{s}}
\def\n{\mb{n}}
\def\z{\mb{z}}

\def\h{\mb{h}}
\def\Lc{\mc{L}}
\def\f{\text{F}}
\def\F{\mb{F}}
\def\Ro{\boldsymbol{\rho}}

\def\A{\text{A}}
\def\A{\text{A}}
\def\T{\text{T}}
\def\R{\text{R}}
\def\Ce{\text{Ce}}
\def\Pr{\textbf{P}}
\newcommand{\cov}{\mathrm{Cov}}

\usepackage{color}

\makeatletter

\newcommand{\Rmnum}[1]{\expandafter\@slowromancap\romannumeral #1@}
\makeatother

\begin{document}

\preprint{------}

\title{Study of statistical properties of {\it hybrid} statistic in coherent multi-detector CBC Search} 
\author{K Haris\footnote{Electronic address: haris@iisertvm.ac.in} and Archana Pai\footnote{Electronic address: archana@iisertvm.ac.in}}
\affiliation{Indian Institute of Science Education and Research Thiruvananthapuram, CET Campus, Trivandrum 695016}
\date{\today}
\begin{abstract}
  In this article, we revisit the problem of coherent multi-detector  
  	search of gravitational wave from compact binary coalescence with Neutron stars and Black Holes using advanced interferometers like LIGO-Virgo. Based on the loss of optimal 
  	multi-detector signal-to-noise ratio (SNR), we construct a {\it hybrid} statistic as a best of maximum-likelihood-ratio(MLR) statistic tuned for face-on and face-off binaries.
  The statistical properties of the {\it hybrid} statistic is studied.
  The performance of this {\it hybrid} statistic is compared with  that of the coherent MLR  statistic for generic inclination angles. Owing to the single synthetic data stream, the {\it hybrid} statistic gives low false alarms compared to the multi-detector MLR statistic and  small fractional loss  in the optimum SNR for a large range of binary inclinations. 
  We have demonstrated that for a  LIGO-Virgo network and binary inclination $\epsilon < 70^\circ$ and $\epsilon > 110^\circ$, the {\it hybrid} statistic captures more than $98\%$
    of network optimum matched filter SNR with low false alarm rate. The Monte-Carlo exercise with two distributions of incoming inclination angles namely, $U[\cos \epsilon]$
    and more realistic distribution proposed in \cite{Schutz:2011tw} are performed with {\it hybrid} statistic and gave $\sim 5 \%$ and $ \sim 7 \%$ higher detection probability respectively
    compared to the two stream multi-detector MLR statistic for a fixed false alarm probability of $10^{-5}$.
\end{abstract}


\maketitle

\section{Introduction}\label{Sec-Intro}

On $14^{th}$ September 2015, the two Advanced LIGO detectors (LIGO-Livingston and LIGO-Hanford)\cite{0264-9381-32-7-074001,Harry:2010zz}  detected gravitational waves (GW)  for the first time from a binary black hole merger event\cite{PhysRevLett.116.061102}. The Advanced Virgo detector will be ready for observation of cosmos very soon\cite{TheVirgo:2014hva,Avirgo}. The Japanese cryogenic detector KAGRA is under construction\cite{PhysRevD.88.043007,0264-9381-29-12-124007} and  proposal for a detector in India namely LIGO-India is in place\cite{LIGOIndia}. Compact binary coalescences (CBC) with Neutron stars (NS) and Black Holes (BH) are one of the most promising GW sources for the Advanced LIGO-Virgo interferometric GW detectors. The Advanced LIGO detectors have a proposed distance reach of $\sim 445 Mpc.$ for binary neutron star (BNS) events
and  are expected to detect few events of BNS inspiral per month\cite{0264-9381-27-17-173001}. Detection of CBC would reveal information about the BH as well as the NS equation of state. We expect many more surprises from nature in the form of GW detections which would emerge into a new, exciting field of GW astronomy in few decades.

The detection of GW in the interferometric data $\x$ is a statistical hypothesis testing problem, where the null hypothesis -- $\x$ {\it is purely noise} $\n$ --  is tested against an alternative  hypothesis --  $\x$ {\it is signal} $\s$ {\it plus the noise} $\n$. The decision is based on construction of a {\it detection statistic} -- a real valued function of $\x$ -- and is compared with a predefined threshold. When this test statistic crosses the threshold, the detection is declared. There are various strategies adopted for setting this threshold. The  common strategy is to fix the {\it false alarm} rate (based on the available prior knowledge of the interferometer noise) and obtain the threshold value for the statistic.

The Neyman - Pearson lemma\cite{nla.cat-vn4318735} says, the likelihood ratio (LR) -- the probability ratio of the data following alternative hypothesis and the null hypothesis --  is the most power full test statistic in case of  simple hypotheses (signal is known). However, in GW detection problem e.g. CBC search or continuous wave search (from a periodic source such as Pulsar), the signal model is known but the parameters are unknown. Here, the alternative hypothesis is a composite hypothesis. There are two approaches to composite hypothesis testing.  The first approach is the maximum likelihood ratio (MLR) approach, where LR is maximized over the signal parameters. In the second approach -- Bayesian approach -- which includes the
astrophysical priors of signal parameters, the LR is marginalized over the signal parameters with a prior distribution.
For high signal-to-noise ratio (SNR), the LR is expected to peak at the actual signal values in the multi-dimensional space of signal parameters. Thus, most of the contribution to the marginalized LR is from the maximum.
Therefore, the MLR statistic  performs equally well as the marginalized statistic in the regime of high SNR.

The coherent multi-detector search of GW combines the incoming GW signal at different interferometers
in a phase coherent way, where the information of the arrival time is incorporated in the phase. The MLR based multi-detector approach for CBC signals is developed in GW literature \cite{Pai:2000zt,Harry:2010fr,Haris:2014fxa}. The non-spinning CBC signal is
a function of 9 parameters; namely masses, source location, amplitude, binary inclination, polarization angle, phase at the time of arrival and time of arrival at the reference detector. The MLR multi-detector statistic obtained by maximizing multi-detector LR over a subset of 4 signal parameters (namely; amplitude, binary inclination, polarization angle and initial phase) was shown to be the sum of
MLR statistic of two synthetic data streams which captures the two GW polarizations in Einsteinian General Relativity. Henceforth, we would refer to this statistic as  generic MLR statistic. In \cite{Williamson:2014wma}, authors investigate the performance of multi-detector MLR statistic devised for face-on/off binaries in the targeted follow-up of short gamma ray burst (SGRB) in the GW window. In \cite{PhysRevD.80.063007}, authors explore Bayesian framework
to address the multi-detector CBC detection problem. The multi-detector
coherent approach for continuous wave search is developed  in \cite{PhysRevD.72.063006} and in \cite{Prix:2009tq}, authors further compare the performance of Bayesian {\it vs} MLR statistic in a specific set of amplitude coordinates given in \cite{PhysRevD.58.063001}.

In this paper, we revisit the MLR based multi-detector CBC statistics. As mentioned above, the generic  multi-detector MLR statistic, $\Lc$ for CBC signal is
a sum of two single streams (synthetic data systems) MLR statistics (Eq.(2.38) of \cite{Harry:2010fr} and Eq.(44) of \cite{Haris:2014fxa}) in
dominant Polarization frame\cite{PhysRevD.72.122002}. In this work, we carefully analyze the statistical properties  of multi-detector MLR statistic for Gaussian noise. Further, we obtain the MLR based statistics specially targeted for the face-on/off binaries which we denote as $\Lc^{0, \pi}$.  This is a single data stream MLR statistic as opposed to the two stream $\Lc$ statistic and gives less {\it false alarm} rate as compared to that of $\Lc$. {\it A careful study of SNRs of $\Lc^{0,\pi}$ indicate that either $\Lc^0$ or $\Lc^\pi$ captures most of the multi-detector optimum SNR for a wide range of inclination angle, $\epsilon$ and polarization angle, $\Psi$.} We have demonstrated that, for $\epsilon < 70^\circ$ and $\epsilon > 110^\circ$, either $\Lc^0$ or $\Lc^\pi$ captures more than $98\%$  of network optimum matched filter SNR.  This is one of the main results of the paper. We further constructed a  {\it hybrid} statistics, $\Lc^{mx} \equiv \max\{\Lc^0, \Lc^\pi\}$ and studied the statistical properties of the same for Gaussian noise. Pertaining to the single stream statistic capturing most of the optimum SNR, the  {\it hybrid} statistic shows less false alarms than the two stream MLR statistic $\Lc$. We perform extensive numerical simulations to
confirm the same. Further, {\it false alarm probability} (FAP) and the {\it detection probability} (DP) obtained from the simulations agrees remarkably well with the proposed analytical expressions.

	In \cite{Williamson:2014wma}, the authors examined the $\Lc^{0,\pi}$
	statistic in the context of targeted follow-up of SGRBs in GW windows. 
	By comparing the inclination angle dependent polarization contributions to the SNR ({\it i.e.} $\cos \epsilon$ and $\frac{1 + \cos^2 \epsilon}{2}$),  authors showed that face-on/off MLR statistic perform better (low false alarms) than the generic multi-detector  MLR statistic for SGRB search. Since the focus was on the follow-up of SGRB in GW window, the observational constraints of jet opening angle restricts the binary
	inclination angle within $30^\circ$ from 0 or $180^\circ$. Thus the study was restricted to the above mentioned range of binary inclinations. On the other hand in this paper, we address generic inclination angles for the non spinning CBC search.     

The paper is divided as follows; In Sec.\ref{MLR}, we review the non-spinning CBC signal, the multi-detector MLR statistic $\Lc$ and  statistical properties of $\Lc$. In Sec.\ref{MLR-faceon-off}, we construct the targeted face-on/off statistic $\Lc^{0,\pi}$ and study their statistical properties. We study the signal SNR in $\Lc^{0,\pi}$ for arbitrary inclination and polarization angles. In Sec.\ref{HybridStatistic}, we propose the  {\it hybrid} statistic $\Lc^{mx}$ and study its statistical properties. In Sec. \ref{Simulations}, we summarize the numerical simulations and discuss the results.

\section{Review of GW CBC coherent multi-detector MLR statistic}
\label{MLR}
In this section, we summarize the earlier works \cite{Pai:2000zt,Harry:2010fr,Haris:2014fxa} on the coherent multi-detector MLR statistic for the detection of non-spinning
CBC signal using  advanced interferometers.

For a network of $I$ interferometric detectors, the incoming GW signal from the non-spinning CBC source in m-th detector
is denoted as $\s_m$. The signal is represented in dominant polarization frame and in frequency domain as given below\cite{Haris:2014fxa},
{\small \bea
\tilde  s_m(f) =  \A ~ \tilde h_0(f) ~e^{i \phi_a} \ll[\ll(\frac{1+\cos^2 \epsilon}{2} \cos  2 \chi + i \cos \epsilon \sin 2 \chi \rr) \f_{+m} \rr.  \nn \\
+  \ll.  \ll(\frac{1+\cos^2 \epsilon}{2} \sin  2 \chi - i \cos \epsilon \cos  2 \chi \rr) \f_{\times m}   \rr],  \label{signal}
\eea}
where the signal parameters are the overall amplitude  $\A$,  initial phase $\phi_a$ (signal phase at the time of arrival in the fiducial reference detector typically coinciding with the Earth's center), the binary inclination angle $\epsilon$, and polarization angle $\Psi = \chi - \delta/4$. The  angle $\delta$ is a function of source direction and distribution of detectors on  Earth, which uniquely  defines the dominant polarization frame of the network for a given source direction. The $\f_m \equiv \f_{+m} +i \f_{\times m} $ is the complex antenna pattern function of the $m^{th}$ detector in dominant polarization frame, which is a function of source location and the multi-detector configuration (location of detectors on Earth's globe)\footnote{Throughout the paper, we express the signal as well as the antenna pattern functions  in dominant polarization frame.}.  $\tilde h_0(f)  \equiv f^{-\frac{7}{6}} e^{i \varphi(f)}$ defines the frequency evolution of the signal, with  the restricted non-spinning 3.5 PN phase  $\varphi(f)$, which is a function of two component masses of the binary and the time of arrival of the signal in the reference detector.
Please note, here we assume that we know the source location (targeted CBC search) and hence the signal $\s_m$ as defined in Eq.(\ref{signal}) is appropriately compensated for the delays in the arrival time. 

For spatially distributed detectors, the noise in individual detector is independent. Thus the network matched filter SNR square,
$\Ro_s^2$ is the sum of squares of SNRs in the individual detectors and is given by,\footnote{The scalar product of $\mb{a}$ and $\mb {b}$ is defined as  $$\langle\mb{a}|\mb{b}\rangle = 4 \Re \int_0^\infty \tilde{\tilde a}(f)~ \tilde b^*(f)~ df,$$ where $\tilde{\tilde{a}}(f) = \tilde a(f)/S_n(f)$ is the double-whitened version of frequency series $\tilde  a(f)$. The $S_n(f)$ is the one sided noise power spectral density(PSD) of a detector.}
\be
\Ro_s^2 = \sum_{m=1}^I \langle \s_m | \s_m\rangle. \label{SNR}
\ee

\subsection{\it Log likelihood ratio}
For interferometers with independent and additive Gaussian noise ($\x = \s +\n$), the network log likelihood ratio(LLR), $\varLambda$ is the sum of LLRs of individual detectors as given below\cite{Pai:2000zt,Harry:2010fr},

\be
2 \varLambda = 2 \sum_{m=1}^I \langle \x_m| \s_m \rangle - \langle  \s_m| \s_m \rangle, \label{LLR1}
\ee
where $\x_m$ is data stream from $m^{th}$ detector. In \cite{Haris:2014fxa}, it is shown that Eq.\eqref{LLR1} is the sum of LLRs of two effective synthetic streams $\z_L$ and $\z_R$ of the network as below.

{\small \bea
2 \varLambda = \ll[2 \Ro_L \langle  \z_L|   \h_0 e^{i  \Phi_L} \rangle - \Ro_L^2 \rr] +\ll[ 2 \boldsymbol{\rho_R} \langle  \z_R|  \h_0 e^{i  \Phi_R} \rangle  - \boldsymbol{\rho_R}^2 \rr]. \label{LLR2}
	\eea}
For a given sky location, the over-whitened synthetic steams, $\tilde {\tilde z}_{L,R}(f)$ are obtained by projecting over-whitened network data  on $(+)$ and $(\times)$ polarizations of the complex network antenna pattern vector in dominant polarization frame as follows,

\be
\tilde {\tilde z}_L(f) \equiv \sum_{m=1}^I \frac{\f_{+ m}}{\|\F^{'}\|} \tilde{\tilde x}_m(f),~~\tilde{\tilde z}_R(f) \equiv \sum_{m=1}^I \frac{\f_{\times m}}{\|\F^{'}\|} \tilde{\tilde x}_m(f).   \label{zLzR}
\ee 

The quantity {\small $\|\F^{'}\|^2 = \sum_{m=1}^I g_m^2 (\f_{+m}^{2} + \f_{\times m}^{2}) $} incorporates the different noise PSDs in different detectors through   $g_m^2 = \langle \h_0|\h_0\rangle$. $g_m$ depicts the difference in individual SNRs of detectors caused by the difference in the noise PSD.

In this notation, the physical parameters $(\A,\phi_a,\epsilon,\Psi)$ are mapped to a new set of parameters $(\Ro_L,\Ro_R,\Phi_L,\Phi_R)$  as shown in Appendix-\ref{appendix-extrinsic}. Similar to the physical parameters, the new set appears
either proportional to amplitude or  phase carrying the extrinsic nature as expected.
From Eq.\eqref{signal},  Eq.\eqref{SNR}  and Eq.\eqref{parametrs}, the multi-detector matched filter SNR square
is distributed in the individual synthetic stream SNRs, $\Ro_{Ls}$ and $\Ro_{Rs}$ as follows.
\be
\Ro^2_s = \Ro^2_{Ls} + \Ro^2_{Rs}. \label{SNR-2}
\ee
where the subscript $s$ refers to the signal.



\subsection{\it Maximization of LLR over extrinsic parameters}

The multi-detector MLR is obtained by  maximizing LLR  over the new parameters  $(\Ro_L,\Ro_R,\Phi_L,\Phi_R)$ and is given in Eq.(44) of \cite{Haris:2014fxa} as below,
{\small \bea
 \Lc \equiv 2 \hat \Lambda = && \underbrace{\langle  \z_L |  \h_0 \rangle^2 + \langle  \z_L |  \h_{\pi/2} \rangle^2}_{\hat \Ro_L^2} \nn \\
&&+\underbrace{\langle  \z_R |  \h_0 \rangle^2 + \langle  \z_R |  \h_{\pi/2} \rangle^2}_{\hat \Ro_R^2}.
\label{MLR1}
\eea}

$\Lc$ can be understood as the sum of power of the synthetic streams $\tilde \z_L$ and $\tilde \z_R$ in two quadratures
$\h_{0,\pi/2}$. In absence of noise, $\Lc$  is equal to the multi-detector matched filter SNR square \cite{Pai:2000zt,Harry:2010fr,Haris:2014fxa} and further
\be
\hat \Ro_{L}|_{\n=0} = \Ro_{Ls} ~~~~~~~~ \hat \Ro_{R}|_{\n=0} = \Ro_{Rs} \,. \label{MLR-nonoise}
\ee

\subsection{\it False alarm and detecton probabilities}
\label{FAP-DP-MLR}
In this section, we summarize statistical properties of $\Lc$. Let $p_0(\Lc)$ be the probability distribution of $\Lc$ in  absence of signal and $p_1(\Lc)$ be the distribution in  presence of signal. For a given threshold $\pounds$, the FAP, $Q_0$ and DP, $Q_d$ are given by,
\be
Q_{0}(\pounds) = \int_{\pounds}^{\infty} p_{0}(\Lc) d\Lc, ~~~~~ Q_{d}(\pounds) = \int_{\pounds}^{\infty} p_{1}(\Lc) d\Lc \,.\label{Q0QPi} 
\ee 



In  absence of signal and for uncorrelated Gaussian noise in detectors the 4 scalar products $\langle  \z_{L,R} |  \h_{0,\pi/2} \rangle$ in $\Lc$ are standard normal variates $\sim \mathcal{N}(0,1)$. Thus, $\Lc$ being a sum square of 4 standard normal variates, it follows a $\chi^2$ distribution with 4 degrees of freedom\cite{Pai:2000zt} {\it i.e.}

\be
p_0(\Lc) =  \frac{\Lc}{4} \exp\ll[ -\Lc/2 \rr]\,. \label{P0-MLR}
\ee 

The FAP becomes,
\be
Q_0(\pounds) = \int_{\pounds}^{\infty} p_0(\Lc) d\Lc  = \ll(1+\frac{\pounds}{2} \rr) \exp\ll[-\pounds/2 \rr]. \label{Q0-MLR}
\ee 

In presence of signal, $\Lc$ is equal to sum of squares of 4 random variables following normal distribution with unit variance and individual means. Using Eq.(\ref{MLR-nonoise}), the sum of squares of the means is equal to $\Ro_s^2$. Thus the distribution of $\Lc$ follows (see Eq.(7.7) in \cite{Pai:2000zt}),
\be
p_1(\Lc) = \frac{1}{2} \frac{\Lc}{\Ro_s} exp\ll[- \frac{\Lc+\Ro_s^2}{2}\rr] I_1(\Ro_s \sqrt{\Lc}),
\ee
 where $I_1$ is the modified Bessel function of second kind with order 1.  In an asymptotic limit $\Ro_s \sqrt{\Lc} \gg 1$, $p_1(\sqrt{\Lc})$ can be approximated by a Gaussian distribution\cite{Pai:2000zt},

\be
p_1(\sqrt{\Lc}) = \frac{1}{2 \pi} \exp\ll[-\frac{(\sqrt{\Lc} - \Ro_s)^2}{2}\rr].
\ee 

The DP can be approximated as,
\be
Q_d(\pounds) = \int_{\pounds}^{\infty} p_1(\Lc)~ d\Lc \approx \frac{1}{2} \textbf{erfc}\ll(\frac{\sqrt{\pounds} - \Ro_s}{\sqrt{2}}\rr), \label{Qd-MLR}
\ee
where $\textbf{erfc}$ is the complimentary error function.

 \section{Maximum likelihood analysis for face-on/off sources}
 \label{MLR-faceon-off}
 In this section, we focus on the two special cases of binaries namely face-on ($\epsilon =0$) and face-off ($\epsilon=180^\circ$) and obtain the MLR statistic.
 
 From Eq.\eqref{signal}, the frequency domain signal  for face-on/off binary is given by, 
{\small 
	\bsube \label{signal-faceon}
	\bea
	\tilde  s^0_m(f) &=&  \A ~ \tilde h_0(f) ~\f^{*}_{m}~e^{i (\phi_a + 2 \chi)},  \\
	\tilde  s^\pi_m(f) &=&  \A ~ \tilde h_0(f)~ \f_{m} ~e^{i (\phi_a - 2 \chi)}.
	\eea
	\esube}
The superscript $0$ and $\pi$ correspond to the face-on and face-off cases respectively. Please note,  the polarization angle is absorbed in the initial phase. Hence both the parameters can not be estimated individually, which gives rise
to reduction in the parameter space by one. From Eq.\eqref{parametrs} the new parameters become,
\bsube \label{faceon-params}
 \bea
 &\Ro_L = \text{A} \|\F_+^{'}\|,~~~~ &\Ro_R =  \frac{\|\F_\times^{'}\|}{\|\F_+^{'}\|}\Ro_L , \label{faceon-params1} \\
 &\Phi_L = \chi + \phi_a,~~~~~~~~~ &\Phi_R = \Phi_L \mp \frac{\pi}{2}. \label{faceon-params2}
 \eea
\esube
 The only difference between the face-on and face-off cases appears in terms of a sign in Eq.\eqref{faceon-params2}. In the expression of $\Phi_R$, the negative sign is for face-on case and positive for the face-off case {\it i.e.} the $\Phi_R$ for face-off gets shifted by $180^\circ$ compared to the $\Phi_R$ in phase-off. Please note, in Eq.\eqref{faceon-params}, $\Ro_R$ and $\Phi_R$ are expressed in terms of $\Ro_L$ and $\Phi_L$. Thus in face-on/off case, LLR statistic is a function of 2 parameters instead of three. Physically, the face-on/off case, amounts to the circular polarization and hence different polarization angles carry no extra information and can not be distinguished from the initial signal phase, $\phi_a$.
 
 If we substitute Eq.\eqref{faceon-params} in Eq.\eqref{LLR2}, the LLR reduces to,  
\be
 2 \Lambda^{0,\pi} = 2 \Ro~ \langle \z^{0,\pi}|\h_0 e^{i\Phi_L} \rangle - \Ro^2, \label{LLR-faceon} 
 \ee
with the new parameter $\Ro \equiv \A \|\F^{'}\|$ and 
{\small \be
  \tilde {\tilde z}^0(f)  \equiv \sum_{m=1}^I \frac{\f_m}{\|\F^{'}\|} \tilde{\tilde x}_m(f),~~~ 
  \tilde {\tilde z}^\pi(f)  \equiv \sum_{m=1}^I \frac{\f^{*}_m}{\|\F^{'}\|} \tilde{\tilde x}_m(f) . \label{Z0Zpi}  
\ee}

Maximization of $\Lambda^{0,\pi}$ over $\Ro$ and $\Phi_L$ gives MLR statistic $\hat \Lambda^{0,\pi}$ as,

{\small \be
	\Lc^{0,\pi} = \langle  \z^{0,\pi} |  \h_0 \rangle^2 + \langle  \z^{0,\pi} |  \h_{\pi/2} \rangle^2 \,. \label{MLR-faceon}
	\ee} 

The statistic is a single data stream statistic of $\z^{0, \pi}$, which is constructed in Eq.(\ref{Z0Zpi}). $\Lc^{0,\pi}$ can be understood as power of $\z^{0,\pi}$ in the two quadratures $\h_{0,\pi/2}$. We expect the multi-detector MLR statistic for face-on/off case to evolve in to such a single stream statistic since signal is proportional to $\F^*$ or $\F$ (see Eq.\eqref{signal-faceon}). We note that the Eq.\eqref{MLR-faceon} is same as
the Eq.(22) of \cite{Williamson:2014wma}.

In  absence of noise the statistics $\Lc^{0,\pi}$ becomes equal to the network matched filer SNR square. {\it i.e.}
\be
\Lc^{0,\pi}|_{\n=0} = \Ro^2_s.
\ee

\subsection{\it False alarm and dismisal probabilities}
\label{FAP-DP-faceon}

In  absence of signal, the scaler products  $\langle  \z^{0,\pi} |  \h_{0} \rangle$ and $\langle  \z^{0,\pi} |  \h_{\pi/2} \rangle$ become standard normal variates. Thus the probability distribution of $\Lc^0$ as well as $\Lc^\pi$ is $\chi^2$ with 2 degrees of freedom. {\it i.e.}

\be
p_0(\Lc^{0,\pi}) = \frac{1}{2} \exp\ll[-\Lc^{0,\pi}/2 \rr]. \label{P0-faceon}
\ee

The FAP with threshold $\pounds$ becomes,

\be
Q^{0,\pi}_0(\pounds) \equiv \int_{\pounds}^{\infty} p_0(\Lc^{0,\pi})~ d\Lc^{0,\pi} = \exp\ll[-\pounds/2 \rr].  \label{Q0-faceon}
\ee

In presence of signal, as in Sec.\ref{FAP-DP-MLR}, $\Lc^{0,\pi}$ is equal to sum of squares of two Gaussian random variables with unit variance and distinct means such that sum of squares of means, $\Lc^{0,\pi}|_{\n=0} = \Ro_s^2$. Then the distribution of $\Lc^{0,\pi}$ is given by Eq.(2.10) of \cite{nla.cat-vn4318735} as,

\be
p_1(\Lc^{0,\pi}) = \frac{1}{2} \exp\ll[-\frac{\Lc^{0,\pi}+\Ro_s^2}{2}\rr] I_0 \ll(\Ro_s \sqrt{\Lc^{0,\pi}}\rr), \label{p1-faceon}
\ee
where $I_0$ is the modified Bessel function of second kind with order 0. 

Similar to $\Lc$, in the asymptotic limit $\Ro_s \sqrt{\Lc^{0,\pi}} \gg 1$, distribution of  $\sqrt{\Lc^{0,\pi}}$ can be approximated by normal distribution
with mean equal to $\Ro_s$ and unit variance. Thus the DP for threshold $\pounds $ can be approximated by $\textbf{erfc}$ function as,

\be
Q^{0,\pi}_d(\pounds) \approx \frac{1}{2} \textbf{erfc}\ll(\frac{\sqrt{\pounds} - \Ro_s}{\sqrt{2}}\rr). \label{Qd-faceon}
\ee

Here we make an important observation that the DP of  $\Lc^{0,\pi}$ in Eq.\eqref{Qd-faceon} is identical with the DP of $\Lc$ in Eq.\eqref{Qd-MLR}
for a fixed multi-detector optimum SNR $\Ro_s$.

We remind the reader that now we have three distinct multi-detector MLR statistics namely, $\Lc$ for unknown inclination angle  and $\Lc^{0,\pi}$ targeting  face-on/off sources. We note the main difference between them is that $\Lc$ is
a 2 data stream statistic while $\Lc^{0,\pi}$ is a single stream statistic (see Eq.\eqref{MLR1} and Eq.\eqref{MLR-faceon}). Thus, from the statistical properties ( see  Eq.\eqref{Q0-MLR} and Eq.\eqref{Q0-faceon}), for a fixed threshold $\pounds$ and given signal SNR $\Ro_s$, the false alarm rate of $\Lc$ would be higher than that of $\Lc^{0, \pi}$.


\begin{figure}
	\centering
	\includegraphics[width=0.5\textwidth]{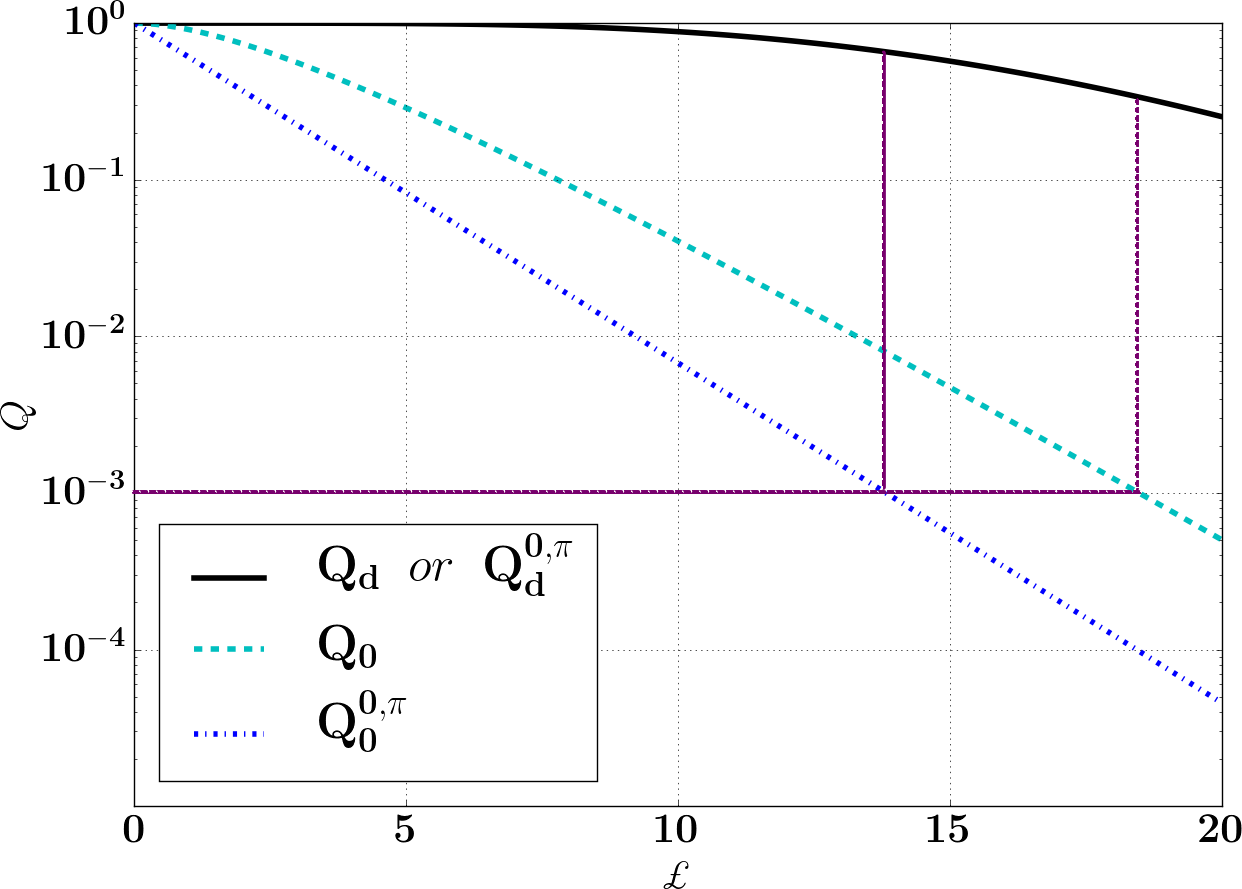}
	\caption{\label{FAP} Variation of FAP and DP of  $\Lc$ and $\Lc^{0,\pi}$ with respect to  the threshold $\pounds$ for $\Ro_s = 4.$ For a fixed FAP of $10^{-3}$, DP of $\Lc^{0,\pi}$ is $0.67$ while that of $\Lc$ is $0.35$.}
\end{figure}

In other words, to achieve the fixed FAP, the threshold for $\Lc^{0,\pi}$ needs to be lowered than that of $\Lc$. Therefore more signal events will cross the threshold when the statistic $\Lc^{0,\pi}$ is used as compared to $\Lc$. This makes $\Lc^{0,\pi}$ a better statistic compared to $\Lc$ in face-on/off case. In Fig.(\ref{FAP}), we have plotted FAP and DP of  $\Lc$ and $\Lc^{0,\pi}$ with respect to  the threshold $\pounds$ for $\Ro_s = 4.$ For example, when we draw a fixed FAP of $10^{-3}$ line, the figure shows that the DP of $\Lc^{0,\pi}$ is $0.67$ while that of $\Lc$ is $0.35$  showing a clear improvement in the DP for $\Lc^{0,\pi}$.



\subsection{\it Performance of $\Lc^{0,\pi}$ for an \text{arbitrary} inclination angle}
\label{performance-faceon}
In this section, we investigate the performance of $\Lc^{0,\pi}$ for an incoming signal from a binary with an arbitrary inclination. First, we study the fractional optimum SNR captured by
$\Lc^{0,\pi}$.

We note in the previous section that $\Lc^{0,\pi}|_{\n=0} = \Ro_s^2$ for face-on/off case. However, if we use the same statistic for an arbitrarily oriented binary
then, the $\z^{0,\pi}$ would capture a fraction of network matched filter SNR and it would drop with increase in $\epsilon$. We denote this fraction by,  $\omega^{0,\pi} \equiv \sqrt{\Lc^{0,\pi}} / \Ro_s$. 

\begin{figure}
	\centering
	\includegraphics[width=0.49\textwidth]{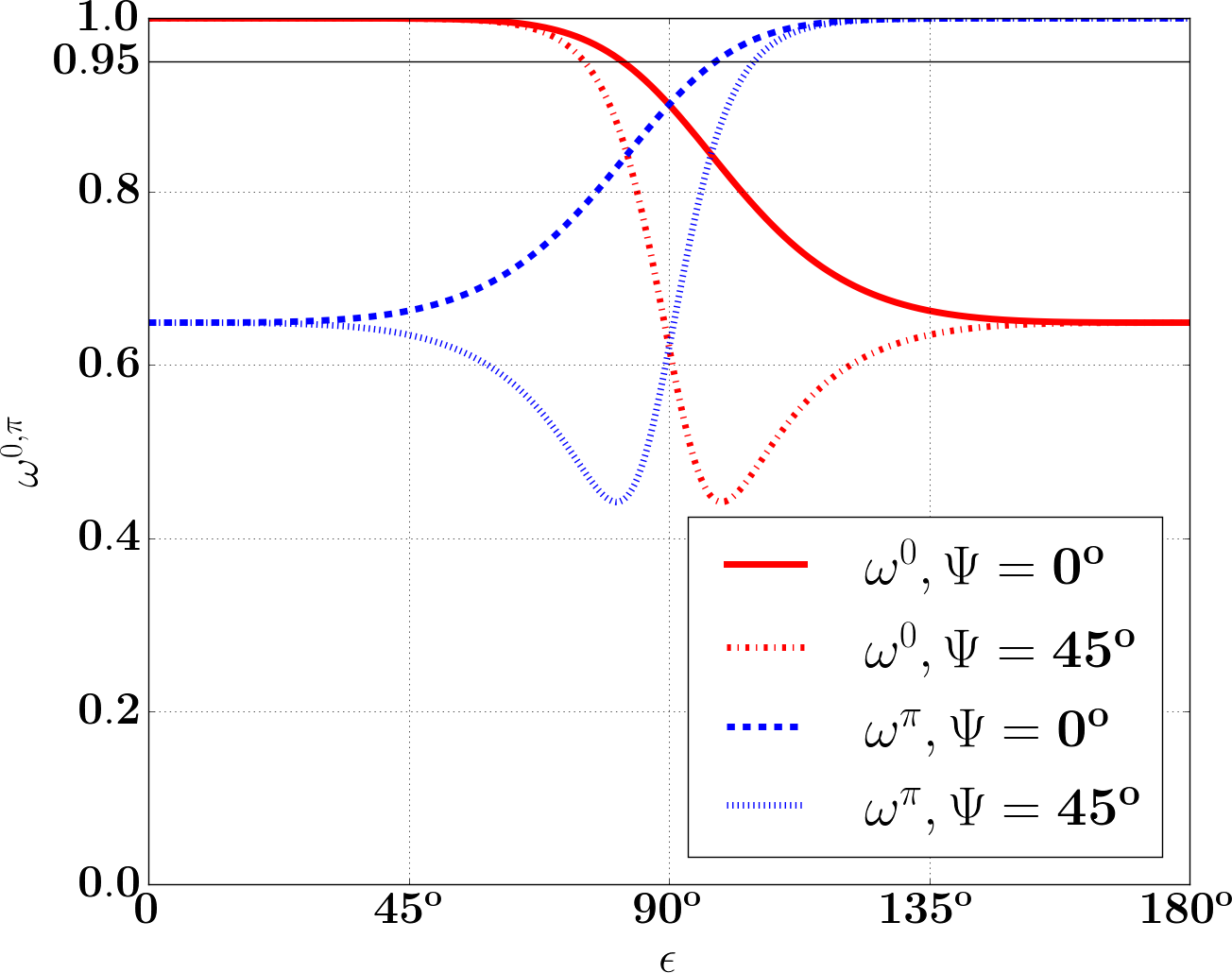}
	\caption{\label{Rho_0Pi_1} Variation of  $\omega^0$ and  $\omega^\pi$ with respect to the inclination angle $\epsilon$ for 2 different values of $\Psi$ in the network LHV. The signal with  SNR, $\Ro_s = 6$ is from $(2 - 10 M_\odot)$ NS-BH system optimally located at $(\theta=140^\circ,\phi=100^\circ)$. We assume "zero-detuning, high power" Advanced LIGO PSD for all detectors\cite{aLIGOSensitivity}. }
\end{figure}

In Appendix-\ref{appendix-L0,LPi_Signal}, we derive the expression for $\omega^{0,\pi}$ and show that for a wide range of $\epsilon$, either $\omega^0$ or $\omega^\pi$ is close one. Specifically, for $\epsilon \leq 70^\circ$, $\omega^0 \approx 1$ and for $110 \leq \epsilon \leq 180^\circ$,  $\omega^\pi \approx 1$. (Please see Appendix-\ref{appendix-L0,LPi_Signal} for details.) This is elaborated in Fig.(\ref{Rho_0Pi_1}). It shows the behavior of $\omega^0$ and $\omega^\pi$ with respect to $\epsilon$  for a network LHV, with Ligo-Livingston (L), Ligo- Hanford (H) and Virgo (V) as the constituent detectors. The signal is from a $(2 - 10  ~M_\odot)$ NS-BH binary located at $(\theta=140^\circ, \phi=100^0)$. We assume fixed multi-detector optimum SNR, $\Ro_s =6$. The plots  are drawn for two different values of polarization angle, $\Psi = 0^\circ$ and $\Psi =45^\circ$. We note that for these values, for a fixed $\Psi$, the fraction $\omega^{0,\pi}$ captures most of the SNR for almost all values of $\epsilon$ except a window of $40^\circ$ centered at $\epsilon = 90^\circ$ (edge on case). Please note that the width of this window has small variation with respect to $\Psi$ as shown in the figure.

\begin{figure*}
	\centering
	\includegraphics[width=0.9\textwidth]{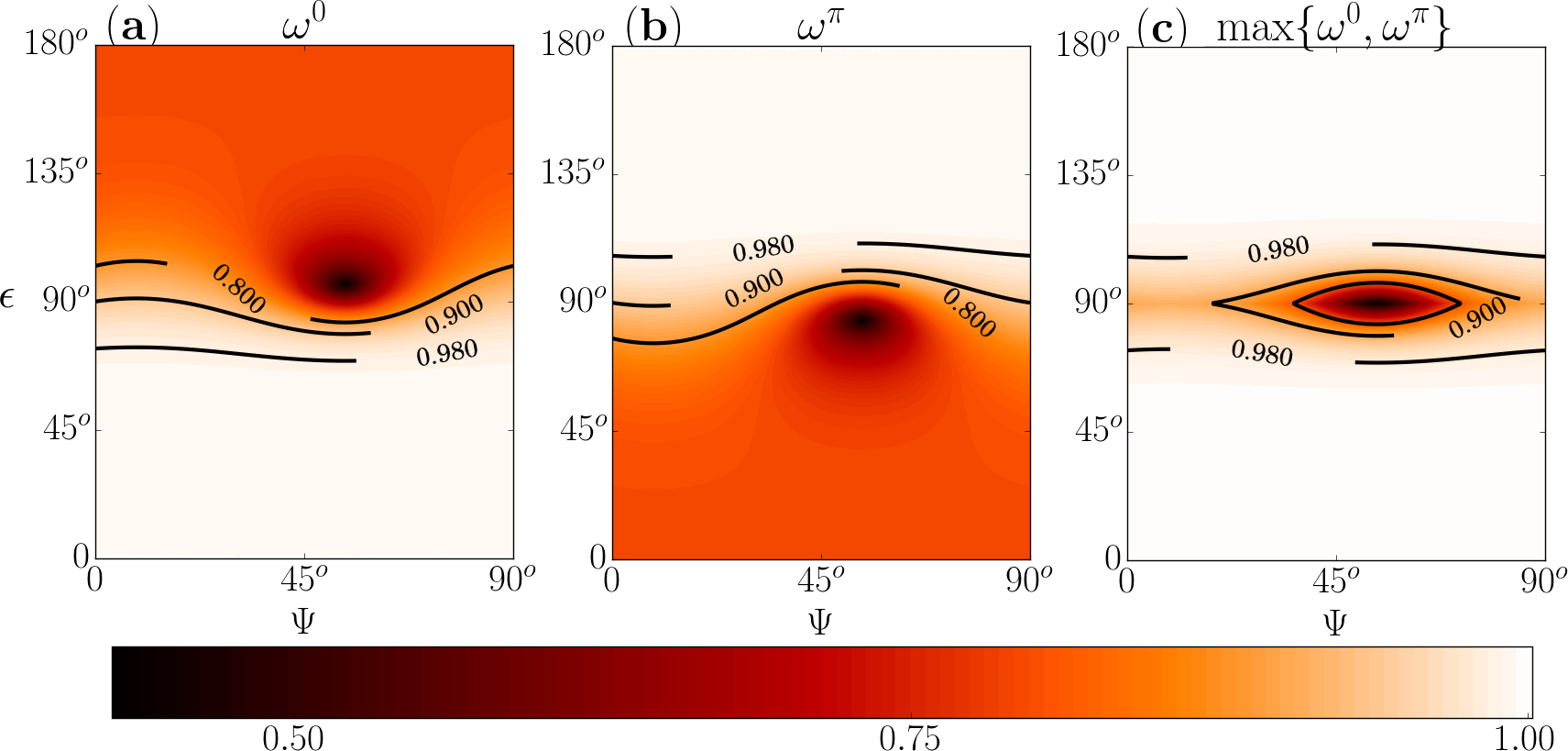}
	\caption{\label{Rho_0Pi_2}Map of $\omega^0$, $\omega^\pi$ and $\max\{\omega^0, \omega^\pi\}$ in $(\epsilon, \Psi)$ plane for a network LHV.
	The signal is  from $(2 - 10 M_\odot)$ NS-BH system optimally located at $(\theta=140^\circ,\phi=100^\circ)$. The multi-detector matched filter SNR, $\Ro_s = 6$. We assume "zero-detuning, high power" Advanced LIGO PSD\cite{aLIGOSensitivity} for all detectors.}
\end{figure*}

 
In Fig.(\ref{Rho_0Pi_2}), we further elaborate the same by drawing the maps of $\omega^{0}$ and  $\omega^{\pi}$ in the $(\epsilon-\psi)$ plane (see panel (a) and (b)). We draw contours of constant $\omega^{0,\pi}$ at values $\omega^{0,\pi} = 0.98, 0.9, 0.8$. It is clear that $\forall~ \epsilon \leq 70^\circ~ \& ~\forall~ \Psi$, $\omega^o \geq 0.98$. Similarly, for $110^\circ \leq \epsilon \leq 180^\circ ~\&~ \forall~ \Psi$, $\omega^\pi \geq 0.98$. A small region of parameters with $70^\circ < \epsilon < 110^\circ$ shows poor response to both $\z^0$  and $\z^\pi$. The
  $\omega^0$ and $\omega^\pi$ are minimum  at the point $\ll(\epsilon =90^\circ, \chi = 45^\circ \rr)$ (Please note, $\chi = \Psi - \delta/4$ as defined in Sec.\ref{MLR}). We expect that the synthetic streams tuned for face-on/off would give poor response to the edge-on binary.  
   Further we note, $\omega^{0}$ and  $\omega^{\pi}$  are complimentary in nature about $\epsilon = 90^0$. In panel (c), we draw the map of $\max\{\omega^0, \omega^\pi\}$. This shows that barring a small region near edge-on, either $\z^0$ or $\z^\pi$ captures large fraction of $\Ro_s$.

In the rest of the section, we comment on the  statistical properties of $\Lc^{0,\pi}$ for an arbitrary inclination. The main difference for an arbitrarily oriented binary from face-on/off case is that $\z^{0,\pi}$ captures a fraction of $\Ro_s$ instead of $\Ro_s$. Thus in presence of signal, the distribution of $\Lc^{0,\pi}$ for any arbitrary $\epsilon$ is same as Eq.\eqref{p1-faceon} with $\Ro_s$ replaced by $\omega^{0,\pi} \Ro_s$ as given below

\be
p_1(\Lc^{0,\pi}) = \frac{1}{2} \exp\ll[-\frac{\Lc^{0,\pi}+ (\omega^{0,\pi} \Ro_s)^2}{2}\rr] I_0 \ll(\omega^{0,\pi} \Ro_s \sqrt{\Lc^{0,\pi}}\rr). \label{p1-faceon_2}
\ee

 Further the DP remains the same as Eq. \eqref{Qd-faceon}, where $\Ro_s$ replaced by $\omega^{0,\pi} \Ro_s$ as given below.

\be
Q^{0,\pi}_d(\pounds) \approx \frac{1}{2} \textbf{erfc}\ll(\frac{\sqrt{\pounds} - \omega^{0,\pi} \Ro_s}{\sqrt{2}}\rr). \label{Qd-faceon_2}
\ee

Since FAP depends only on noise model and construction of statistic, the FAP 
of $\Lc^{0,\pi}$ for an arbitrary inclination is same as Eq.\eqref{Q0-faceon}.


As we discussed earlier, $\Lc^0$ captures more than $98\%$ of $\Ro_s$ for $\epsilon \leq 70^\circ$ while $\Lc^\pi$ captures more than $98\%$ of $\Ro_s$ for $\epsilon \geq 110^\circ$. Further, the
Fig.(\ref{Rho_0Pi_2}) shows the complimentary behavior of the two statistics $\Lc^0$ and $\Lc^\pi$. In addition, both $\Lc^0$ and $\Lc^\pi$ are constructed out of a
single synthetic stream as opposed to the $\Lc$ statistic (two streams). This motivates us to construct a  {\it hybrid} statistic out of $\Lc^0$ and $\Lc^\pi$ which would capture most of the multi-detector SNR for a large range of binary inclinations for the CBC search.



\section{Proposal of Hybrid Statistic $\Lc^{mx}$}
\label{HybridStatistic}


In this section we propose a {\it hybrid} statistic as $\Lc^{mx} \equiv \max\{\Lc^0,\Lc^\pi\}$ and study its statistical properties.

In  absence of signal,  both $\Lc^0$ and $\Lc^\pi$ follows $\chi^2$ distribution with 2 degrees of freedom (see Eq.\eqref{P0-faceon}). Let  $\Pr(\Lc^0,\Lc^\pi)$ be the
joint probability distribution of $\Lc^0$ and $\Lc^\pi$  then, the probability distribution of $\Lc^{mx}$ can be written down as
\be
p_0(\Lc^{mx}) = 2 \int_0^{\Lc^{mx}} \Pr(\Lc^0=\Lc^{mx}, \Lc^\pi)~ d\Lc^\pi . \label{p0-L-1}
\ee

Please note, here $\Lc^{0}$ and $\Lc^{\pi}$ have non zero covariance. i.e, they are not independent of each other. 

In Eq.\eqref{L0LPi-2}, $\Lc^{0,\pi}$ is expressed in terms of $\z_{L,R}$ as,

{\small \bea
\Lc^{0,\pi} &=& \ll( \frac{\|\F^{'}_+\|}{\|\F^{'}\|} ~ \langle \z_L|\h_0\rangle \pm \frac{\|\F^{'}_\times\|}{\|\F^{'}\|}~ \langle \z_R|\h_\pi/2 \rangle \rr)^2 \nn \\
&+& \ll( \frac{\|\F^{'}_+\|}{\|\F^{'}\|}~ \langle \z_L|\h_\pi/2 \rangle  \mp \frac{\|\F^{'}_\times\|}{\|\F^{'}\|} ~\langle \z_R|\h_0 \rangle \rr)^2 . \label{Hybrid-1}
\eea}

In absence of signal, each of $\langle \z_{L,R}|\h_{0,\pi}\rangle$ follows independent Gaussian distribution with zero mean and unit variance. This ensures the terms inside the two brackets in Eq.\eqref{Hybrid-1}  follow Gaussian distribution with zero mean and unit variance. This implies,
\be
\Lc^0  \equiv  n_1^2 + n_2^2 , ~~~~~ \Lc^\pi  \equiv  n_3^3 + n_4^2, \label{Hybrid-2}
\ee
 such that $n_{1,2,3,4}$ as standard normal variates with 
 {\small \bea
 \cov(n_1,n_3)  &=& \cov(n_2,n_4) = \frac{ \|\F^{'}_+\|^2 -  \|\F^{'}_\times\|^2}{ \|\F^{'}\|^2} \equiv c , \nn \\
  \cov(n_1,n_2) &=& \cov(n_3,n_4) = 0.				 				
\eea}

Then the joint distribution of $\sqrt{\Lc^{0}}$ and $\sqrt{\Lc^{\pi}}$ is a 2-dimensional  generalized Rayleigh distribution and is given by Eq.(2.1) of \cite{blumenson1963} as,
{\small \bea
\Pr(\sqrt{\Lc^0}, \sqrt{\Lc^\pi}) = &&\frac{\sqrt{\Lc^0 \Lc^\pi}}{c}~  e^{-\frac{\Lc^0 +\Lc^\pi}{2 \ll( 1- c^2 \rr)}} \nn \\ 
&& I_0 \ll(\frac{c}{1-c^2} \sqrt{\Lc^0 \Lc^\pi}\rr) . \label{JoinDist-1}
\eea}

This implies,
{\small \bea
	\Pr(\Lc^0,\Lc^\pi)  &=& \frac{1}{4 \sqrt{\Lc^0 \Lc^\pi}} \Pr(\sqrt{\Lc^0}, \sqrt{Lc^\pi}) \nn \\
	 &=& \frac{1}{4~c} ~e^{-\frac{\Lc^0 +\Lc^\pi}{2 \ll( 1- c^2 \rr)}}~  I_0 \ll(\frac{c}{1-c^2} \sqrt{\Lc^0 \Lc^\pi}\rr). \label{JoinDist-2}
	\eea}

Substitution of Eq.\eqref{JoinDist-2}  in Eq.\eqref{p0-L-1} gives the distribution of $\Lc^{mx}$ in absence of signal as, 

{\small \bea
p_0(\Lc^{mx} &)&  = \frac{1}{2~c}~  e^{-\frac{\Lc^{mx} }{2 \ll( 1- c^2 \rr)}} \nn \\ 
&{}&\int_0^{\Lc^{mx}}  e^ {-\frac{\Lc^\pi }{2 \ll( 1- c^2 \rr)}} ~I_0 \ll(\frac{c}{1-c^2} \sqrt{\Lc^{mx} \Lc^\pi}\rr)~ d\Lc^\pi. \label{p0-L-2}
\eea}

In presence of signal, for high multi-detector matched filter SNR $\Ro_s$,  as discussed in Sec.\ref{performance-faceon},  the distribution of $\sqrt{\Lc^{0,\pi}}$ can be approximated by Gaussian distribution with mean $\omega^{0,\pi} \Ro_s$ and unit variance. But for high $\Ro_s$, $\Lc^{mx}=\Lc^{0}$ in the region $0^\circ \leq \epsilon \leq 70^\circ$ and $\Lc^{mx} = \Lc^{\pi}$ in the region $110^\circ \leq \epsilon \leq 180^\circ$. Thus the distribution of $\Lc^{mx}$ in the presence of signal can be approximated as,

  \[
  p_1(\Lc^{mx}) \approx
  \begin{cases}
    p_1(\Lc^0),   & 0 \leq \epsilon < 70^\circ , \\
    p_1(\Lc^\pi), & 110^\circ < \epsilon \leq 180^\circ \,.
  \end{cases}
  \]

The FAP and DP of $\Lc^{mx}$ can be obtained by numerically integrating $p_0(\Lc^{mx})$ and $p_1(\Lc^{mx})$.

In the next section, we carry out numerical simulations to study the statistical properties of $\Lc,\Lc^{0,\pi}$ and the {\it hybrid} statistic $\Lc^{mx}$. Further, we study the performance of all 4 statistics in terms of the Receiver Operator Characteristic (ROC) curve for various signal configurations.

\section{Simulations and Discussion}
\label{Simulations}

\begin{figure*}
	\centering
	\subfloat{\includegraphics[width=1.\textwidth]{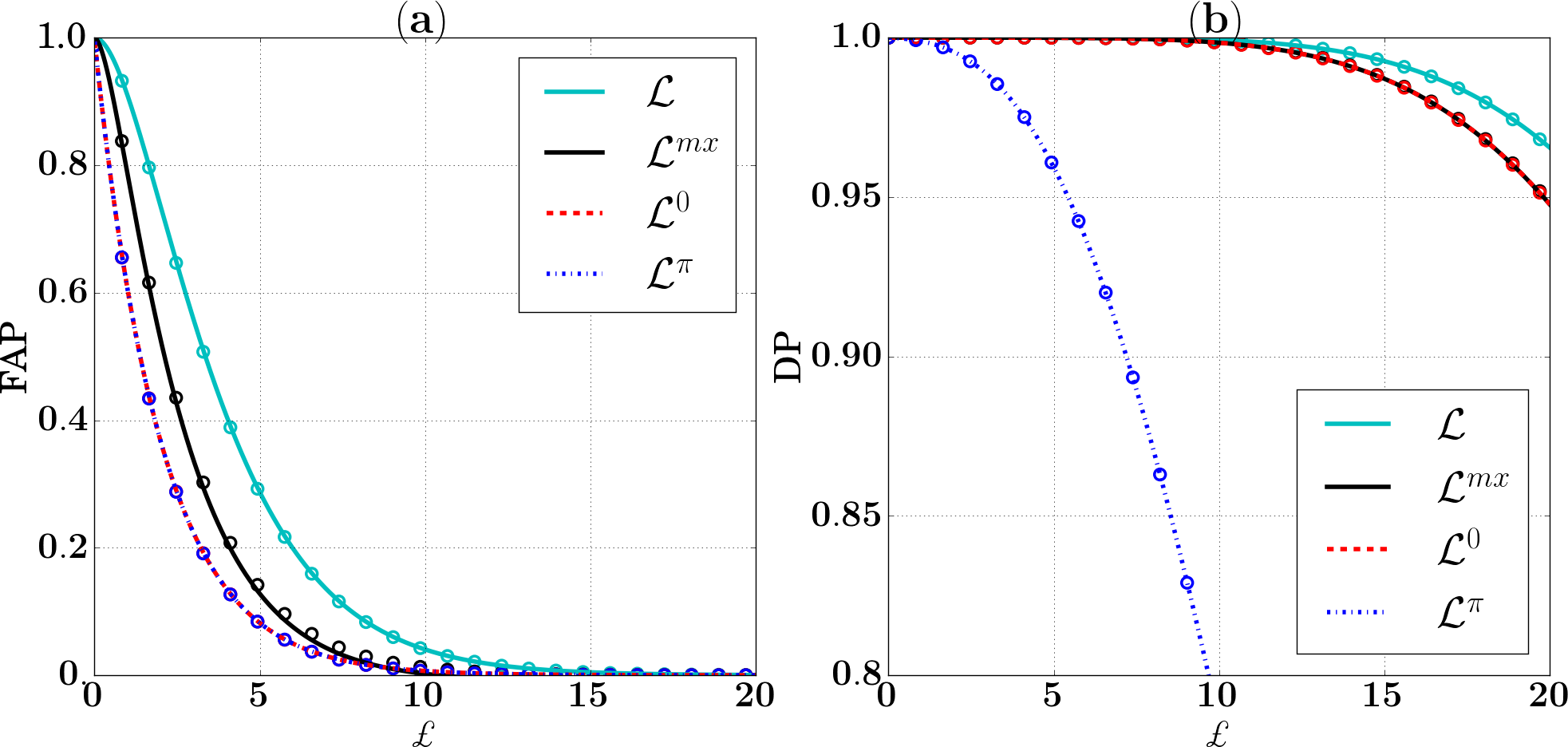}}
	\caption{\label{FAPDP} (a) Variation in FAP of different statistics with respect to the threshold $\pounds$. (b)  Variation in DP of different statistics with respect to the threshold $\pounds$ for  signal from  $(2 -10 M_\odot)$ non spinning NS-BH binary system with $\Ro_s=6$ optimally located at $(\theta=140^\circ,\phi=100^\circ)$ with arbitrary $\epsilon = 45^\circ$ and $\Psi = 45^\circ$. The curves are generated from the theory and circles are from simulations. We assume "zero-detuning, high power" Advanced LIGO PSD\cite{aLIGOSensitivity} for all detectors.}
\end{figure*}

In this section, we carry out numerical simulations for a three detector network LHV. All the detectors are assumed to have Gaussian, random noise with the noise PSD following "zero-detuning, high power"
Advanced LIGO noise curve\cite{aLIGOSensitivity}. The GW signal from non spinning NS-BH ($2 - 10~M_\odot$) binary system
is injected with SNR $\Ro_s=6$. We assume that the masses are fixed and known for this comparison study. Of course, in real situation, the masses are unknown and	then one needs to place templates in mass space and perform the search.  We know that a template based search increases the false alarms. However, this applies to the search based on both {\it hybrid} statistic, $\Lc^{mx}$ and the MLR statistic, $\Lc$ and further owing to a single stream, we expect to get less false alarms for {\it hybrid} statistic as compared to the MLR statistic. As mentioned in Sec.\ref{Sec-Intro}, based on simple arguments
in gravitational wave follow-up of short Gamma Ray Bursts of IPN triggers, in  \cite{Williamson:2014wma} authors  used a face on/off tuned MLR statistic (single stream) for nearly on-axis GRBs. This was targeted search
with templates in mass parameter space in LIGO-Virgo data. They did show a similar improvement in the false alarm rates
compared to generic MLR statistic as we got  in the fixed masses simulations  of {\it hybrid} described below.

The simulation results are as follows. First we compare the theoretical and numerically evaluated FAP and DP for  all the 4 statistics $\Lc$, $\Lc^{mx}$, $\Lc^0$ and $\Lc^\pi$ and then the performance of the {\it hybrid} statistic is compared with the generic MLR statistic, $\Lc$. This performance is quantified by drawing
the ROC plots, i.e, the plot between FAP and DP. In all the plots, The $\Lc$ statistic is represented by cyan(solid) line, $\Lc^{mx}$ by black(solid) line,
$\Lc^0$ by red(dash) line and $\Lc^\pi$ by blue(dash-dot) line.

\subsection{\it Comparison of analytical and numerical FAP and DP}
In Sec.\ref{HybridStatistic}, we obtain the analytical expression for distributions, of $\Lc^{mx}$ in  presence and absence of signal {\it i.e.} $p_{0}(\Lc^{mx})$ and  $p_{1}(\Lc^{mx})$. Theoretical FAP and DP for different thresholds is computed by integrating $p_{0,1}(\Lc^{mx})$. Here, we compare the theoretical FAP and DP with those obtained by numerical simulations.

We generate the network data with $2 \times 10^6$  noise realizations with a fixed signal from NS-BH system located at $(\theta=140^\circ,\phi=100^\circ)$
(one of the best location for LHV network based on the joint antenna power response) with $\epsilon = \psi = 45^\circ$. For each noise realization, all the four $\Lc$, $\Lc^{mx}$, $\Lc^0$, $\Lc^\pi$ statistics are computed. For a given threshold $\pounds$, we count the number of times each of the statistics crosses the threshold value  when the data contains only the noise (gives FAP) as well as when the data contains signal plus noise (gives DP).

In Fig.(\ref{FAPDP}), panel (a) represents the FAP {\it vs} $\pounds$  for all the 4 statistics. The open circles denote  FAP computed through simulations as detailed above whereas the continuous lines denote the theoretically obtained FAP.
  We observe a remarkable agreement of the analytical result with the numerical simulation. The main result derived and understood from  panel (a) of Fig.(\ref{FAPDP}), is
  the difference in the FAP values corresponding to a  given threshold for various statistic. Owing to two data streams, $\Lc$ gives the maximum FAP amongst all the four. Since
  $\Lc^{0,\pi}$ are constructed out of a single synthetic stream, the FAP of both is identical as well as the least amongst the four. Since the {\it hybrid} statistic is
  constructed out $\Lc^{0,\pi}$, its FAP is slightly higher than that of $\Lc^{0,\pi}$.

The panel (b) of Fig.(\ref{FAPDP}), represents the DP {\it vs} $\pounds$ for all the 4 statistics. The open circles denote the DP from simulations  whereas the continuous lines denote the theoretical DP. Since the DP for all the 4 statistics depends on the
fractional optimal SNR captured by the individual statistic, the DP of $\Lc$ is maximum as it captures $\Ro_s$ in no noise case. For the signal with $\epsilon= 45^\circ$, $\Lc^{mx}$ is $\Lc^0$ most of the time, thus the DP of $\Lc^{mx}$ and $\Lc^0$ overlap. The statistic $\Lc^\pi$ captures a small fraction of the $\Ro_s$ (see Fig.(\ref{Rho_0Pi_2})) and hence shows the least DP. Once again, we see remarkable agreement of numerically computed DP with the analytically integrated DP for all the statistics.
  
\subsection{\it Performance of {\it hybrid} statistic for a single injection}
In this subsection, we study the performance of $\Lc^{mx}$ against rest of the statistics, more importantly the generic multi-detector MLR statistic $\Lc$.

\begin{figure*}
	\centering
	\subfloat{\includegraphics[width=1.\textwidth]{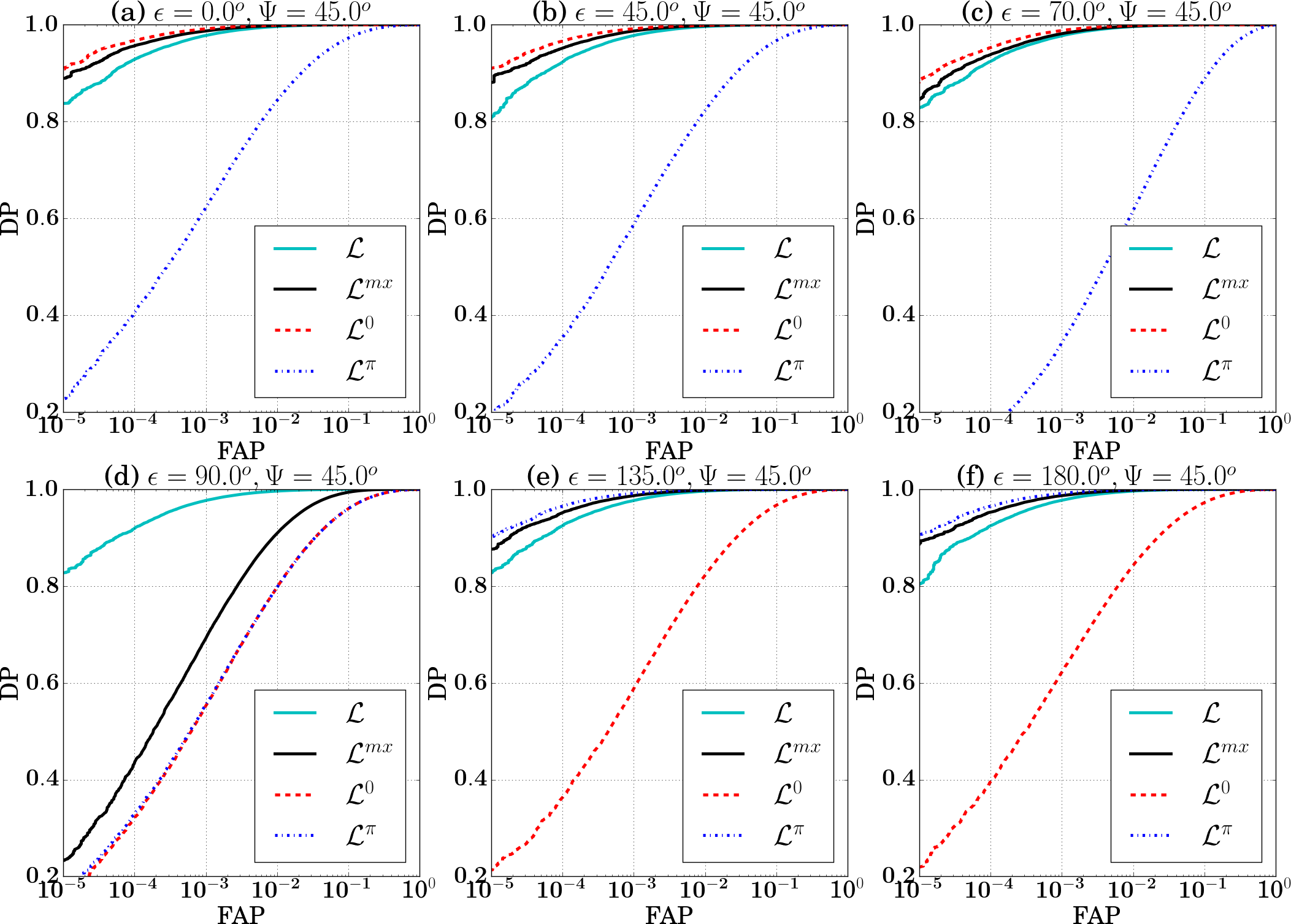}}
	\caption{\label{Roc1}ROC plots of the 4 statistics corresponding to a network LHV for fixed injections with different values of $\epsilon$. The signal with SNR, $\Ro_s = 6$ is from $(2 - 10 M_\odot)$ NS-BH system optimally located at $(\theta=140^\circ,\phi=100^\circ)$ with an arbitrary polarization angle $\psi = 45^\circ$. We assume "zero-detuning, high power" Advanced LIGO PSD\cite{aLIGOSensitivity} for all detectors.}
\end{figure*}

We generate the network data with $2 \times 10^6$ noise realizations and a fixed signal from NS-BH system optimally
located at $(\theta=140^\circ,\phi=100^\circ)$ with an arbitrary $\psi = 45^\circ$ but varying binary inclination. We select 6 binary inclination angles
namely $\epsilon = 0^\circ, 45^\circ, 70^\circ, 90^\circ, 135^\circ, 180^\circ$ and obtain the ROC curve numerically as shown in panels (a), (b),
(c), (d), (e), (f)  of Fig.(\ref{Roc1}) respectively. We summarize the results as below.

{\it For $\epsilon =0~OR~\pi$ case}, $\Lc^{mx}$ (being optimized for the face-on/off case) is expected to perform better than the generic MLR statistic.
Panel (a) and (e) shows the same. As discussed earlier, this improvement is primarily due to the reduction in the FAP of $\Lc^{mx}$. 
For a fixed FAP of $10^{-5}$, the subsequent improvement in the DP is $6\%$ which translates in an increase in the detection rate of $6\%$.

{\it For $\epsilon = 45^{\circ}~OR~135^\circ$ case}, (symmetrical located from $0$ and $\pi$ cases respectively) as seen in
panel (b) and (e), the improvement in ROC of $\Lc^{mx}$ compared to that of $\Lc$ is similar. This improvement is due to the drop in
FAP of $\Lc^{mx}$. As shown in Appendix-\ref{appendix-L0,LPi_Signal}, at $\epsilon = 45^\circ$,
the $\Lc^{mx}$ captures all the optimum SNR. Thus the improvement in DP remains close to $6\%$ similar to the face-on/off case.

Following the above argument, as $\epsilon$ approaches the edge-on case, the fractional SNR captured in $\Lc^{mx}$ reduces. Thus ROC of $\Lc^{mx}$
starts approaching that of ROC of $\Lc$ as seen in panel (c). Here, the improvement of $\Lc^{mx}$ is $2 \%$ over the $\Lc$.

{\it For $\epsilon = 90^{\circ}$ case}, the fractional optimal SNR captured by $\Lc^{mx}$ is very small as $\Lc^{mx}$ is optimized for face-on/off case.
Thus, the MLR statistic performs better than the $\Lc^{mx}$ at the edge-on case as shown in panel (d) of Fig.(\ref{Rho_0Pi_2}).

\begin{figure*}
	\centering
	\subfloat{\includegraphics[width=1.\textwidth]{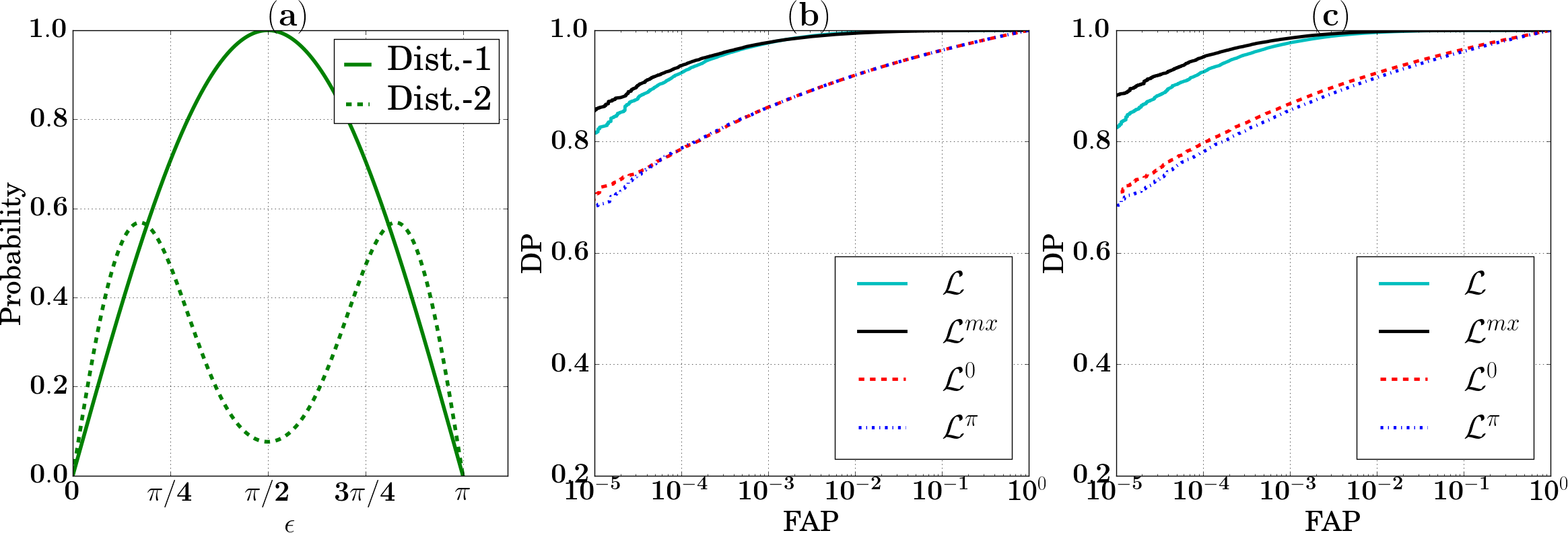}}
	\caption{\label{Roc2} Panel (a) is the plot of  two sampling distributions of $\epsilon$. Panel (b) gives  ROC plots for 4 different statistics  corresponding to a network LHV when the injected signal's inclination angle, $\epsilon$  drawn from {\bf Dist-1} and panel (c) gives ROC plots for injections with $\epsilon$ drawn from {\bf Dist-2}. In both cases  sky location, and polarization angle are sampled uniformly. The injections are with SNR, $\Ro_s = 6$ and are from $(2 - 10 M_\odot)$ NS-BH system. We assume "zero-detuning, high power" Advanced LIGO PSD\cite{aLIGOSensitivity} for all detectors.}
\end{figure*}

\subsection{\it Performance of the {\it hybrid} statistic for injections sampled from a distribution}

In this simulation, we generate the network data with $2 \times 10^6$ noise realizations and  signals from NS-BH system with masses $(2 - 10 M_\odot)$ and multi-detector SNR $\rho_s =6$. We randomly draw the binary inclination angle $\epsilon$, polarization angle $\Psi$ and source location $(\theta, \phi)$  from a given distribution. We perform this exercise for two distinct distributions, {\bf Dist-1} and {\bf Dist-2} of inclination angle $\epsilon$. In both cases, $\cos \theta$, $\phi$ and $\Psi$ are sampled uniformly from the intervals $[-1,1]$, $[0^\circ, 360^\circ]$ and $[0^\circ, 90^\circ]$ respectively.

The {\bf Dist-1}, draws $\cos(\epsilon)$ uniformly from [-1,1] and is denoted by a green (solid) line in panel (a) of Fig.(\ref{Roc2}).
As seen, in the figure, the population of random samples drawn from this distribution contains  more of edge-on sources than that of face-on.

In {\bf Dist-2}, the $\epsilon$ follows the distribution proposed in Eq.(28) of \cite{Schutz:2011tw}  [see green (dashed) line in panel (a) of Fig.(\ref{Roc2})].
\be
\Pr (\epsilon) = 0.076076 \ll( 1+ 6 \cos^2 \epsilon + \cos^4 \epsilon \rr)^{3/2} \sin \epsilon .
\ee

The {\bf Dist-2} is a realistic distribution of $\epsilon$, where the   SNR  information is folded in the distribution along with the geometric prior. Since we know that the edge-on sources have less SNR than face-on sources, we expect to see less number of edge-on systems than face-on. As a result, there
would be a dip in the curve (dashed line) with respect to the {\bf Dist -1} (solid line).
  
In Fig.(\ref{Roc2}), panels (b) and (c) summarize the results in terms of the  ROC curves using {\bf Dist -1} and {\bf Dist -2} respectively.
The ROC curve summarizes the  performance of MLR statistic compared to the {\it hybrid} statistic averaged over all the source locations and the polarizations.

Panel (b) shows that for {\bf Dist -1}, the average performance of $\Lc^{mx}$ is  better than that of $\Lc$ in spite of more number of sources located around edge-on. Quantitatively, DP improves by $\sim 5\%$ for FAP $10^{-5}$.  

However, panel (c) shows more realistic performance as we expect the inclination angle distribution to be more realistic in this case. We
note that for {\bf Dist - 2}, the {\it hybrid} statistic performs much better than the MLR statistic $\Lc$. Quantitatively, for an FAP of $10^{-5}$,
$\Lc^{mx}$ improves the DP by $~7\%$ over $\Lc$. 

\subsection{Conclusion and future directions}

In this article, we revisit the problem CBC GW search with a multi-detector network with advanced interferometers like LIGO-Virgo in coherent approach.
We show that the {\it hybrid} statistic constructed from  two statistics namely; coherent MLR statistic tuned for face-on and face-off binaries captures most of multi-detector optimum SNR for a large fraction of the binary inclination angles except a small window centered around the edge-on case. The statistical properties of this {\it hybrid} statistic is studied in detail. The performance of this {\it hybrid} statistic is compared with that of the coherent MLR  statistic for generic inclination angles. 
Being constructed from the single synthetic data stream, the {\it hybrid} 
statistic gives low false alarms compared to the two  streams generic multi-detector MLR statistic and  very small fractional loss in the optimum SNR for a large range of binary inclinations. 

We have demonstrated the performance by using the noise model as Gaussian with  "zero-detuning, high power" Advanced LIGO PSD\cite{aLIGOSensitivity} in LHV network for the NS-BH system of masses ($2-10 M_\odot$) for a fixed SNR 
of 6. The ROC curves are used as a tool for this demonstration. The simulations are performed for two cases. 

{\it Case 1}: The source is optimally located in the LHV network
and oriented with various binary inclination angles. The ROC curves show 
the {\it hybrid} statistic performs better than the generic MLR statistic for all the
inclination angles less then $70^\circ$ and greater than $110^\circ$. The improvement in each of them corresponds to two factors. First, the {\it hybrid} statistic captures most of the optimum SNR for a large region of inclination and polarization parameter space. Thus, we do not loose much in the DP for a given multi-detector matched filter SNR $\Ro_s$. Further, by construction the  {\it hybrid} statistic is out of a single stream. Thus, the FAP of the {\it hybrid} statistic is better than that of the two stream generic MLR statistic (of course it is slightly worst than the pure single streams $\Lc^0$ and $\Lc^\pi$ ). 


{\it Case 2}: The source location as well as the orientation and polarization 
are sampled from a distribution. The source location is sampled uniformly from the sky sphere. The polarization angle follows a uniform distribution. The inclination angles are drawn from two distributions namely $U[\cos(\epsilon)]$
and more realistic distribution proposed in \cite{Schutz:2011tw} . The ROC curve shows that the performance of {\it hybrid} statistic gives an improvement of $\sim 5 \%$ and $ \sim 7 \%$ respectively in DP 
compared to the generic multi-detector MLR statistic for a fixed FAP of  $10^{-5}$.

In \cite{Williamson:2014wma}, authors applied a similar statistic for the SGRB follow-up search
for very small inclination angles. However, this study and its performance
in Gaussian noise clearly shows that the ${\it hybrid}$ statistic would give
better performance for wide range of inclination angles barring a small window
around the edge-on case. Since, we expect that the source population would
have bias towards the face-on/off cases due to the relative difference in the
SNRs, this statistic would play a crucial role in the multi-detector CBC search in the advanced era.

We plan to apply this for the S6 noise of the science run of LIGO detectors and test the performance of the statistic for generic inclination angles. We also plan to extend the study for a larger network, which  includes LIGO-India and KAGRA.

\section{Acknowledgment}
The authors availed the 128 cores computing facility established by the MPG-DST Max Planck Partner Group  at IISER TVM. The authors would like to thank S. Fairhurst for reviewing the draft and for useful comments. This document has been assigned LIGO laboratory document number LIGO-P1500221.

\appendix

\section{Relation between  $\{\A,\phi_a,\epsilon,\Psi\}$ and $\{\Ro_L,\Ro_R,\Phi_L,\Phi_R\}$}
\label{appendix-extrinsic}
The new parameters, $\{\Ro_L,\Ro_R,\Phi_L,\Phi_R\}$ are related to the physical parameters, $\{\A,\phi_a,\epsilon,\Psi\}$ as below,
{\small
	\bea
  \Ro_L e^{i \Phi_L} &=&   \A \|\F_+^{'}\| e^{i \phi_a} \ll[\frac{1+\cos^2 \epsilon }{2} \cos 2 \chi + i \cos \epsilon \sin 2 \chi \rr] , \nn  \\
  \Ro_R e^{i \Phi_R} &=&   \A \|\F_\times^{'}\| e^{i \phi_a} \ll[\frac{1+\cos^2 \epsilon }{2} \sin 2 \chi - i \cos \epsilon \cos 2 \chi \rr] . \nn \\ \label{parametrs}
  \eea}
The absolute values and the phases of the above equations are $\{\Ro_L,\Ro_R,\Phi_L,\Phi_R\}$ and explicitly given in Eq.(B1) of \cite{Haris:2014fxa}.

\section{ $\Lc^{0,\pi}$ in absence of noise}
\label{appendix-L0,LPi_Signal}
In this section we derive the expression for fraction of multi-detector matched filter SNR captured by  the statistics, $\Lc^0$ and $\Lc^\pi$. 

From Eq.\eqref{zLzR} and Eq.\eqref{Z0Zpi},  $\z^{0,\pi}$ can be re-expressed in terms of $\z_{L,R}$ as below.
\be
\tilde{\tilde z}^{0,\pi}(f) = \frac{\|\F^{'}_+\|}{\|\F^{'}\|} ~\tilde{\tilde z}_L(f) 
\pm i  \frac{\|\F^{'}_\times\|}{\|\F^{'}\|}~ \tilde{\tilde z}_R(f), \label{Z0Pi_3} 
\ee
where $(+)$ corresponds to $\z^0$ and $(-)$ corresponds to $\z^\pi$.
By substituting back in Eq.\eqref{MLR-faceon}, $\Lc^{0,\pi}$ can be expanded in terms of the four terms $\langle \z_{L,R}|\h_{0,\pi}\rangle$.

{\small \bea
	\Lc^{0,\pi} &=& \ll( \frac{\|\F^{'}_+\|}{\|\F^{'}\|} ~ \langle \z_L|\h_0\rangle \pm \frac{\|\F^{'}_\times\|}{\|\F^{'}\|}~ \langle \z_R|\h_{\pi/2} \rangle \rr)^2 \nn \\
	&+& \ll( \frac{\|\F^{'}_+\|}{\|\F^{'}\|}~ \langle \z_L|\h_{\pi/2} \rangle  \mp \frac{\|\F^{'}_\times\|}{\|\F^{'}\|} ~\langle \z_R|\h_0 \rangle \rr)^2 . \label{L0LPi-2}
	\eea}

Using Eq.\eqref{signal} and Eq.\eqref{parametrs}, the scalar products in the above equation in the absence of noise can be written as,
{\small \bea
\langle \z_L| \h_0\rangle|_{\n=0} = \Re \ll[\Ro_{L} e^{i \Phi_{L}} \rr] &\,,& \langle \z_L| \h_{\pi/2}\rangle|_{\n=0} = - \Im \ll[\Ro_{L} e^{i \Phi_{L}} \rr] , \nn \\
\langle \z_R| \h_0\rangle|_{\n=0} = \Re \ll[\Ro_{R} e^{i \Phi_{R}} \rr] &\,,& \langle \z_R| \h_{\pi/2}\rangle|_{\n=0} = - \Im \ll[\Ro_{R} e^{i \Phi_{R}}\rr] . \label{z-scalrproducts} \nn \\
\eea}
Substituting in Eq.\eqref{L0LPi-2} give,
 
 {\small \be
 	\Lc^{0,\pi}|_{\n=0} = \ll| \frac{\|\F^{'}_+\|}{\|\F^{'}\|}~\Ro_{Ls}~ e^{i \Phi_{Ls}}\pm i~ ~\frac{\|\F^{'}_\times\|}{\|\F^{'}\|}~ \Ro_{Rs} ~e^{i \Phi_{Rs}} \rr|^2. \label{Eq-Rho_0Pi-2}
 	\ee}
We further expand Eq.\eqref{Eq-Rho_0Pi-2} in terms of the physical parameters to obtain the explicite dependence on $\epsilon$ for fixed SNR case.
 \begin{widetext}
  {\small \bea
  	\Lc^{0,\pi}|_{\n=0}  &=& \frac{\text{A}^2}{\|\F^{'}\|^2} \ll|  \|\F^{'}_+\|^2 \ll(\frac{1+\cos^2 \epsilon }{2} \cos 2 \chi + i \cos \epsilon \sin 2 \chi \rr) \pm~ i   \|\F^{'}_\times\|^2  \ll( \frac{1+\cos^2 \epsilon}{2} \sin 2 \chi - i \cos \epsilon \cos 2 \chi \rr) \rr|^2  \nn \\
     &=& \frac{\text{A}^2}{\|\F^{'}\|^2} \ll[ \text{T}_1~ \ll(\frac{1+\cos^2 \epsilon}{2}\rr)^2   + \text{T}_2~  \cos^2\epsilon \pm \text{T}_3~ \frac{1+\cos^2 \epsilon}{2} \cos \epsilon \rr]~ , 
      \label{Eq-Rho_0Pi-3}
     \eea}
 \end{widetext}
 where the three terms $T_1, T_2, T_3$ are defined as below.
 {\small \bea
 \text{T}_1 &\equiv& \|\F^{'}_+\|^4 \cos^2 2 \chi + \|\F^{'}_\times \|^4 \sin^2 2 \chi , \nn \\
 \text{T}_2 &\equiv& \|\F^{'}_+\|^4 \sin^2 2 \chi + \|\F^{'}_\times \|^4 \cos^2 2 \chi , \nn \\
 \text{T}_3 &\equiv& 2 \|\F^{'}_+\|^2  \|\F^{'}_\times\|^2 ~. \label{T1T2T3}
 \eea}
 To obtain a fixed multi-detector matched filter SNR $\Ro_s$, face-one binaries should be kept at larger distance than the edge-on binaries. This is because, the face-on binaries carry more polarization power than the edge-on.
 This reflects in the derived amplitude, $\A$ in Eq.\eqref{Eq-Rho_0Pi-3} as below,
 \be
 \A^2 = \frac{\Ro_s^2}{\text{R}_1 \ll(\frac{1+\cos^2 \epsilon}{2}\rr)^2 + \text{R}_2 \cos^2 \epsilon }, \label{A-fixedSNR}
 \ee
 with
 \bea
 \text{R}_1 &=&  \|\F^{'}_+\|^2 \cos^2 2 \chi + \|\F^{'}_\times \|^2 \sin^2 2 \chi ,  \nn \\
 \text{R}_2 &=&  \|\F^{'}_+\|^2 \sin^2 2 \chi + \|\F^{'}_\times \|^2 \cos^2 2 \chi~. \label{R1R2}
 \eea
 Substitution  of  Eq.\eqref{A-fixedSNR} in Eq.\eqref{Eq-Rho_0Pi-3} gives the fraction, $\omega^{0,\pi}$ of $\Ro_s$ captured by $\Lc^{0,\pi}$ statistic  in the absence of noise. 
{ \small \bea
 (\omega^{0,\pi})^2  &\equiv&  \frac{\Lc^{0,\pi}|_{\n=0}}{\Ro^2_s} \nn \\&=& \frac{ \text{T}_1~ \ll(\frac{1+\cos^2 \epsilon}{2}\rr)^2   + \text{T}_2~  \cos^2\epsilon \pm \text{T}_3~ \frac{1+\cos^2 \epsilon}{2} \cos \epsilon }{{\|\F'\|^2} \ll[\text{R}_1 \ll(\frac{1+\cos^2 \epsilon}{2}\rr)^2 + \text{R}_2 \cos^2 \epsilon \rr]}. \nn \\ \label{Eq-Rho_0Pi-4}
 \eea}

Please note that for the face-on/off case, $\omega^{0,\pi} = 1$. However, as we seen in Fig.(\ref{Rho_0Pi_2}), the $\omega^0$ to drop as the signal $\epsilon$ increases from 0 and similarly, we expect $\omega^\pi$ to drop as the signal $\epsilon$ drops from $\pi$.

In Eq.\eqref{Eq-Rho_0Pi-4}, the inclination angle $\epsilon$ appear in terms of $\cos \epsilon$ and $\frac{1+ \cos^2 \epsilon}{2}$. If we expand $\cos \epsilon$  about $\epsilon =0 $ up to fourth order, then
\bea
\cos \epsilon &\approx& 1- \frac{\epsilon^2}{2} + \frac{\epsilon^4}{24} \equiv \Ce, \nn \\
\frac{1+\cos^2 \epsilon}{2} &\approx&  1- \frac{\epsilon^2}{2} + \frac{4 \epsilon^4}{24} . \label{epsiExpansion}
\eea
Substitution of Eq.\eqref{epsiExpansion} in the expression for $\omega^0$ gives,
{\small \bea
(\omega^0)^2 &=& \ll. \ll(1+ \frac{\T_1+ \T_3 / 2}{4~ \|\F^{'} \|^4~ \Ce}~ \epsilon^4 \rr) \rr/ \ll(1 + \frac{\R_1}{4~ \|\F^{'} \|^2~ \Ce }~  \epsilon^4 \rr) \nn \\
&\approx& \ll( 1+ \frac{\T_1+ \T_3 / 2}{4~ \|\F^{'} \|^4~ \Ce}~ \epsilon^4 \rr) \ll(1 - \frac{\R_1}{4~ \|\F^{'} \|^2~ \Ce }~  \epsilon^4 \rr)  \nn \\
&\approx& 1 + \frac{\epsilon^4}{\Ce~ \|\F'\|^4} \ll( \T_1+ \T_3 /2 - \|\F'\|^2 \R_1 \rr), \label{omega01}
\eea}
Here we make use of the  identities, {\small $\T_1+\T_2+\T_3 = \|\F'\|^4$} and {\small $\R_1 +\R_2 = \|\F'\|^2$} from Eq.\eqref{T1T2T3} and Eq.\eqref{R1R2}. Again from Eq.\eqref{T1T2T3} and Eq.\eqref{R1R2},

\be
 \T_1+ \T_3 /2 - \|\F'\|^2 \R_1 = 0. 
\ee 
This implies,
\be 
\omega^0 =1.\label{omega02}
\ee

The $4^{th}$ order approximation of $\cos \epsilon$ given in Eq.\eqref{epsiExpansion} is valid for a wide range of $\epsilon$. For $\epsilon \leq 70^\circ$, the error in this approximation is less than $1.5 \%$. Thus we can safely assume $\omega^0 \approx 1$  for $0^\circ \leq \epsilon \leq 70^\circ $. For example, Table \ref{Table1} gives the minimum values of $\omega^o$ over $\Psi$ for various $\epsilon$ in a network LHV for  signal from $(2 - 10 M_\odot)$ NS-BH system  located at $(\theta=140^\circ,\phi=100^\circ)$.

\begin{table}[h]
	\begin{tabular}{|l|c|c|c|r|}
		\hline
		Binary Inclination, $\epsilon$ & $50^\circ$ & $60^\circ$ & $70^\circ$ & $80^\circ $ \\ \hline
		Minimum value of  $\omega^0$ & $0.999$ & $0.995$ & $0.98$ & $0.87$
		\\ \hline  
	\end{tabular}
	\caption{Minimum value of $\omega^o$ over $\Psi$ for various $\epsilon$ in a network LHV. The signal is from $(2 - 10 M_\odot)$ NS-BH system  located at $(\theta=140^\circ,\phi=100^\circ)$. We assume "zero-detuning, high power" Advanced LIGO PSD\cite{aLIGOSensitivity} for all detectors.  }
	\label{Table1}
\end{table}

Similarly, by expanding $\cos \epsilon$ about $\pi$, it can be easily shown that   for $110^\circ \leq \epsilon \leq 180^\circ $,  $\omega^\pi \approx 1$. Fig.(\ref{Rho_0Pi_1}) and  Fig.(\ref{Rho_0Pi_2}) justifies the above claim.

For edge-on binaries, at $\chi = 45^\circ$ (Please note, $\chi = \Psi - \delta/4$ as defined in Sec.\ref{MLR}), from Eq.\eqref{Eq-Rho_0Pi-4} the SNR fraction captured by $\Lc^{0,\pi}$ becomes,
\be
\omega^{0,\pi} = \frac{\|\F_\times'\|}{\|\F'\|}.
\ee 
In other words, at this point $\Ro_L$ vanishes and whole network SNR is accumulated in SNR, $\Ro_R$ of sub-dominant stream $\z_R$.  Since by construction of dominant polarization frame, $\|\F_\times'\|$ is less than $\|\F_+'\|$, this result in a minimum $\omega^{0,\pi}$.


\bibliography{paper}
\end{document}